\documentclass[aps,prl,reprint,superscriptaddress]{revtex4-1}

\usepackage[english]{babel}
\usepackage[utf8x]{inputenc}
\usepackage[T1]{fontenc}

\usepackage{amsmath}
\usepackage{amssymb}
\usepackage{graphicx}
\usepackage{slashed}
\usepackage{subfigure}
\usepackage{rotating}
\usepackage{longtable}
\usepackage{gensymb}
\usepackage[colorinlistoftodos]{todonotes}

\begin{document}
\title{Supersymmetry with Dark Matter is still natural}

\author{Melissa van Beekveld}
\email{mcbeekveld@gmail.com}
\affiliation{Institute for Mathematics, Astrophysics and Particle Physics, Faculty of Science, Mailbox 79, Radboud University Nijmegen, P.O. Box 9010, NL-6500 GL Nijmegen, The Netherlands}

\author{Wim Beenakker}
\email{w.beenakker@science.ru.nl}
\affiliation{Institute for Mathematics, Astrophysics and Particle Physics, Faculty of Science, Mailbox 79, Radboud University Nijmegen, P.O. Box 9010, NL-6500 GL Nijmegen, The Netherlands}
\affiliation{Institute  of  Physics,  University  of  Amsterdam,  Science  Park  904,  1018  XE  Amsterdam,  The Netherlands}
\author{Sascha Caron}
\email{scaron@cern.ch}
\affiliation{Institute for Mathematics, Astrophysics and Particle Physics, Faculty of Science, Mailbox 79, Radboud University Nijmegen, P.O. Box 9010, NL-6500 GL Nijmegen, The Netherlands}
\affiliation{Nikhef, Science Park, Amsterdam, The Netherlands}
\author{Ruud Peeters}
\email{r.j.c.peeters@rug.nl}
\affiliation{Van Swinderen Institute for Particle Physics and Gravity, University of Groningen, Nijenborgh 4, 9747 AG Groningen, The Netherlands}
\author{Roberto Ruiz de Austri}
\email{rruiz@ific.uv.es}
\affiliation{Instituto de Física Corpuscular, IFIC-UV/CSIC, Valencia, Spain}


\date{\today}

\begin{abstract}
\noindent We identify the parameter regions of the phenomenological  minimal supersymmetric standard model (pMSSM) with the minimal possible fine-tuning.  We show that the fine-tuning of the pMSSM is not large, nor under pressure by LHC searches. Low sbottom, stop and gluino masses turn out to be less relevant for low fine-tuning than commonly assumed. We show a link between low fine-tuning and the dark matter relic density.  Fine-tuning arguments point to models with a dark matter candidate yielding the correct dark matter relic density: a bino-higgsino particle with a mass of $35-155$ ~GeV. Some of these candidates are compatible with recent hints seen in astrophysics experiments such as Fermi-LAT and AMS-02. We argue that upcoming direct search experiments, such as XENON1T, will test all of the most natural solutions in the next few years due to the sensitivity of these experiments on the spin-dependent WIMP-nucleon cross section.
\end{abstract}
\maketitle
\noindent It is expected that the Standard Model of particle physics (SM) is only an effective theory that needs to be complemented at higher energies.  The problem of extending the SM arises in the high sensitivity of the Higgs potential to the mass scale of new physics. If this scale largely exceeds the electroweak scale we generally have the so-called fine-tuning (FT) problem: a huge degree of cancellation is needed between the tree-level mass and the independent quantum corrections to match the measured Higgs boson mass \cite{Hall2012}.
For many years supersymmetry (SUSY) \cite{Martin:1997ns} with particles at the TeV scale was regarded to be the most natural solution to the FT problem due to a cancellation of fermionic and bosonic contributions to the quantum corrections \cite{KAUL198219, WITTEN1981513}. Furthermore, SUSY is motivated as providing the most general space-time symmetry, a unification of coupling constants and a starting point to solve the shortcomings of the SM.

\noindent In addition, R-parity conserving SUSY provides through the lightest neutralino ($\tilde{\chi}^0_1$) one of the best weakly interacting massive particle (WIMP) candidates for dark matter (DM). Within the $\Lambda$CDM model, Planck measurements of the cosmic microwave background yield a value for the dark matter relic density: $ \Omega_{\rm DM, Planck}h^2 = 0.1186 \pm 0.0011$ \cite{Planck:2015}.

\noindent Due to the null results at the various collider and DM experiments, there is a growing current of opinion that SUSY is just another beautiful idea that didn’t pan out. The main argument is that SUSY particles already need to be so heavy, that SUSY itself requires a significant amount of FT to reproduce the electroweak scale correctly, making the theory unnatural independent of the FT measure used \cite{Papucci:2011wy, PhysRevD.91.075005, Drees:2015aeo}.   One must realize that this statement is framework dependent, e.g. particular GUT scale models such as CMSSM or gauge-mediated SUSY are indeed fine-tuned \cite{Baer:2012mv, Baer:2013gva, Casas:2016xnl, Baer:2014ica} . However, this is no longer true if we consider a less constrained SUSY extension of the SM \cite{Ross:2017kjc, CahillRowley:2012rv,Cici:2016oqr, Baer:2013vpa,Buckley:2016kvr}.

\noindent In this paper we re-evaluate the FT of SUSY by looking at the minimal SUSY extension of the SM (MSSM), restricted to the phenomenologically most relevant soft SUSY breaking parameters. The phenomenology of a whole class of SUSY GUT models is embedded in this framework. Our conclusions are therefore applicable to a whole range of SUSY extensions of the SM.  By algorithmically minimizing the FT in the SUSY parameter space, we look for solutions with the lowest possible FT in the MSSM framework. We check whether these solutions are compatible with current phenomenological constraints and provide a good candidate for DM.  Since in the MSSM the mass of the lightest Higgs boson is linked directly to the Z-boson mass,  we will use the Z-boson mass to quantify the amount of FT.  

\section{Fine-tuning measure via the Z mass}
\noindent A generic SUSY theory has two relevant energy scales: a high-scale one, at which the SUSY breaking takes place, and a low-scale one, usually indicated by $M_{\rm S}$, at which the resulting supersymmetric particle (sparticle) spectrum is situated. Within the MSSM, the mass of the $Z$-boson ($m_Z$) can be expressed in SUSY parameters via minimization of the one-loop Coleman-Weinberg effective potential  \cite{Baer:2012cf, Coleman:1973jx}:  
\begin{equation}
\label{eq:mz}
\frac{m_Z^2}{2} = \frac{m_{H_d}^2 + \Sigma_d^d - (m_{H_u}^2 + \Sigma_u^u)\tan^2\beta}{\tan^2\beta-1} - \mu^2,
\end{equation}
where $m_{H_d}$ and $m_{H_u}$ are the soft SUSY-breaking Higgs masses, $\mu$ the SUSY version of the SM Higgs-mass parameter and $\tan\beta$ the ratio of the vacuum expectation values of the two neutral Higgs fields. The two effective potential terms $\Sigma_u^u$ and $\Sigma_d^d$ denote the one-loop corrections \cite{PhysRevD.87.115028dd}. All terms in expression (\ref{eq:mz}) are evaluated at the energy scale $M_{\rm S}$, which we take to be the geometric average of the two stop masses.\\
\noindent Several measures can be used to quantify the degree of FT \cite{Ellis:1986yg, BARBIERI198863, Kitano200558, PhysRevD.73.095004, PhysRevLett.109.161802}. These measures regard a model to be fine-tuned if the size of a term on the right-hand-side of eq. (\ref{eq:mz}) is much larger than $m_Z^2$ itself, or if $m_Z$ is sensitive to a small variation of one the pMSSM parameters. Since we try to find the minimal possible FT of the pMSSM, we use a measure of FT that is unambiguous and model independent, i.e. independent of unknown high-scale parameter choices or the mechanism by which sparticles acquire their masses. To this end, we employ the directly observable low-scale sparticle spectrum to define the FT.
The foregoing arguments lead us to use the so called electroweak measure $\Delta_{\rm EW}$ \cite{PhysRevLett.109.161802, PhysRevD.87.115028dd} as a measure of the FT:
\begin{equation}
\label{eq:FT}
{\rm FT} \equiv \Delta_{\rm EW} \equiv \max_i \left\lvert\frac{C_i}{m_Z^2/2}\right\rvert,
\end{equation}
where the $C_i$ are defined as:
\begin{align*}
C_{m_{H_d}} &= \frac{m_{H_d}^2}{\tan^2\beta-1},\hspace{1em} C_{m_{H_u}} =  \frac{-m_{H_u}^2\tan^2\beta}{\tan^2\beta-1},\hspace{1em}  C_{\mu} = -\mu^2 \\
C_{\Sigma_d^d} &= \frac{\max(\Sigma_d^d)}{\tan^2\beta-1},\hspace{1em} C_{\Sigma_u^u} = \frac{-\max(\Sigma_u^u)\tan^2\beta}{\tan^2\beta-1}.
\end{align*}
\noindent For $\Sigma_u^u$ and $\Sigma_d^d$ the contributions originating from different particles are considered separately and the maximum contribution is used to define $C_{\Sigma_d^d}$ and $C_{\Sigma_u^u}$ \cite{PhysRevD.87.115028dd}. \\
To obtain a low FT, we generally expect that the sparticles that dominate the FT measure have a mass that is not too far away from the EW scale. Please note that $\Delta_{\rm EW}$ is a FT measure that gives rise to conservative conclusions. A given sparticle spectrum, being agnostic about how it actually came about, will give rise to a unique value of the FT, regardless of any renormalization group trajectory that should have been used to translate between the high-scale underlying theory and that particular sparticle spectrum. For all models with low FT ($<10$), we explicitly evaluated also the sensitivity of $m_Z$ to small variations of the pMSSM parameters. All models with low FT are not found to be more sensitive to these variations than the FT would imply, showing that there is no intrinsic FT in the terms of eq. \ref{eq:FT}.

\section{Scanning the phenomenological MSSM}
\noindent The MSSM has 105 non-SM Lagrangian parameters, including complex phases. One can reduce this number to 19 by using phenomenologically motivated constraints. These constraints comprise of taking degenerate first and second generation squark and slepton masses, setting to zero all trilinear couplings of the first and second generation sfermions, not allowing for new sources of CP violation and demanding for minimal flavor violation. This defines the so-called phenomenological MSSM (pMSSM) \cite{Djouadi:1998di}. 
\noindent For our exploration of the pMSSM we use SUSPECT \cite{Djouadi:2002ze} as spectrum generator. MicrOMEGAs 4.2.5 \cite{Barducci:2016pcb} is used to compute  $\Omega_{\rm DM} h^2$, the velocity weighted DM annihilation cross section ($\langle \sigma v\rangle$) and the spin-dependent and spin-independent WIMP-nucleon scattering cross sections ($\sigma_{\rm SD}$ and $\sigma_{\rm SI}$). The FT is computed using an in-house code, which is checked for consistency with predictions from ISASUGRA from ISAJET 7.85 \cite{Paige:2003mg}. We have checked that our FT calculation gives the same FT for the spectrum resulting from SUSPECT as for the spectrum resulting from FeynHiggs \cite{Bahl:2016brp, Hahn:2013ria,Frank:2006yh,Degrassi:2002fi, Heinemeyer:1998yj}, irrespective of the fact that SUSPECT systematically gives higher values for the Higgs boson mass than FeynHiggs. 
\noindent In order to efficiently explore the parameter space, we begin by choosing the pMSSM model parameters randomly according to the uniform distribution in the box indicated in table 2 of ref. \cite{Caron:2016hib}. We sample all sparticle mass parameters up to 4 TeV, except for the first and second generation squark and slepton mass parameters, which are fixed at 3.5 TeV since their contribution to the FT is small. In an iterative procedure the minimal FT points of the foregoing iteration are used as seeds to sample new model points, where a truncated multi-dimensional Gaussian distribution is used as width around each parameter of the seed to sample new points \cite{1232326}.

\subsection{Limits applied to the model points}
\noindent The following limits are applied to the model points:
\begin{itemize}
\item LEP limits on the masses of the chargino ($m_{\tilde{\chi}^{\pm}_1}~>~103.5$~GeV) and sleptons ($m_{\tilde{l}}~>~90$~GeV) \cite{LEP:working}.
\item Constraints on the invisible and total width of the $Z$-boson,  $\Gamma_{Z, {\rm inv}}~=~499.0~\pm~1.5$~MeV and $\Gamma_{Z}~=~2.4952~\pm~0.0023$~GeV respectively, obtained from $Z$-pole measurements at LEP \cite{Carena:2003aj}.
\item The LHC measurements of the Higgs boson mass \cite{Aad:2014aba, Cao:2014efa}. On top of this we account for a theoretical SUSY uncertainty of 3 GeV, selecting models with a Higgs boson within the mass range of $122$ GeV $\leq m_{h_0} \leq 128$ GeV. We have checked that the Higgs mass output of SUSPECT and FeynHiggs are both in this range.
\item An upper bound of the muon anomalous magnetic dipole moment $\Delta(g-2)_{\mu} < 40\times 10^{-10}$, taking into account the fact that the SM prediction lies well outside the experimentally obtained value: $(24.9\pm 6.3)\times 10^{-10}$ \cite{Roberts:2010cj}.
\item Measurements of the $B/D$-meson branching fractions Br$(B_{(s)}^0~\rightarrow~\mu^+\mu^-)$~\cite{Aaij:2013aka}, Br$(\bar{B}~\rightarrow~X_s~\gamma)$~\cite{Misiak:2015xwa, Czakon:2015exa}, Br($B^+~\rightarrow~\tau^+~\nu_{\tau}$)~\cite{Kronenbitter:2015kls}, Br$(D_s^+~\rightarrow~\mu^+~\nu_{\mu})$~\cite{Widhalm:2007ws} and Br($D^+_s~\rightarrow~\tau^+~\nu_{\tau})$~\cite{Onyisi:2009th}.
\item  Results of Higgs searches at LEP, the Tevatron and the LHC as implemented in HiggsBounds 4.3.1 \cite{Bechtle:2015pma}. 
\item A determination of the exclusion of a model point using SUSY-AI.  SUSY-AI is a machine learning tool, trained with ATLAS data, which is able to exclude model points in the 19 dimensional pMSSM parameter space \cite{Caron:2016hib, Barr:2016sho}.  The corresponding training data are documented in ref. \cite{Aad:2015baa} (8 TeV and 20.3 fb$^{-1}$) and ref. \cite{Barr:2016inz} (13 TeV and 3.2 fb$^{-1}$). 
\item Constraints on the WIMP-nucleus scattering cross section from LUX and PICO, using LUXcalc \cite{Savage:2015xta} updated with the 2016 results from LUX \cite{Akerib:2016vxi} and the 2017 limits from PICO \cite{Amole:2017dex} .
\end{itemize}

\noindent We allow for a multi-component DM and therefore the LUX and PICO limits have to be rescaled by $\frac{\Omega_{\rm DM}}{\Omega_{\rm DM, Planck}}$ if the dark matter relic abundance is less than $\Omega_{\rm DM, Planck}$. As in ref. \cite{Aad:2015baa}, we reject models that are excluded by LUX or PICO with more than $3\sigma$ to account for the form factor uncertainties.  For all other observables we require the value as calculated from the model parameters to lie within the $2\sigma$ interval around the experimentally obtained value. 

\section{Results}

\begin{figure}[t]
	\centering
	\includegraphics[width=0.5\textwidth]{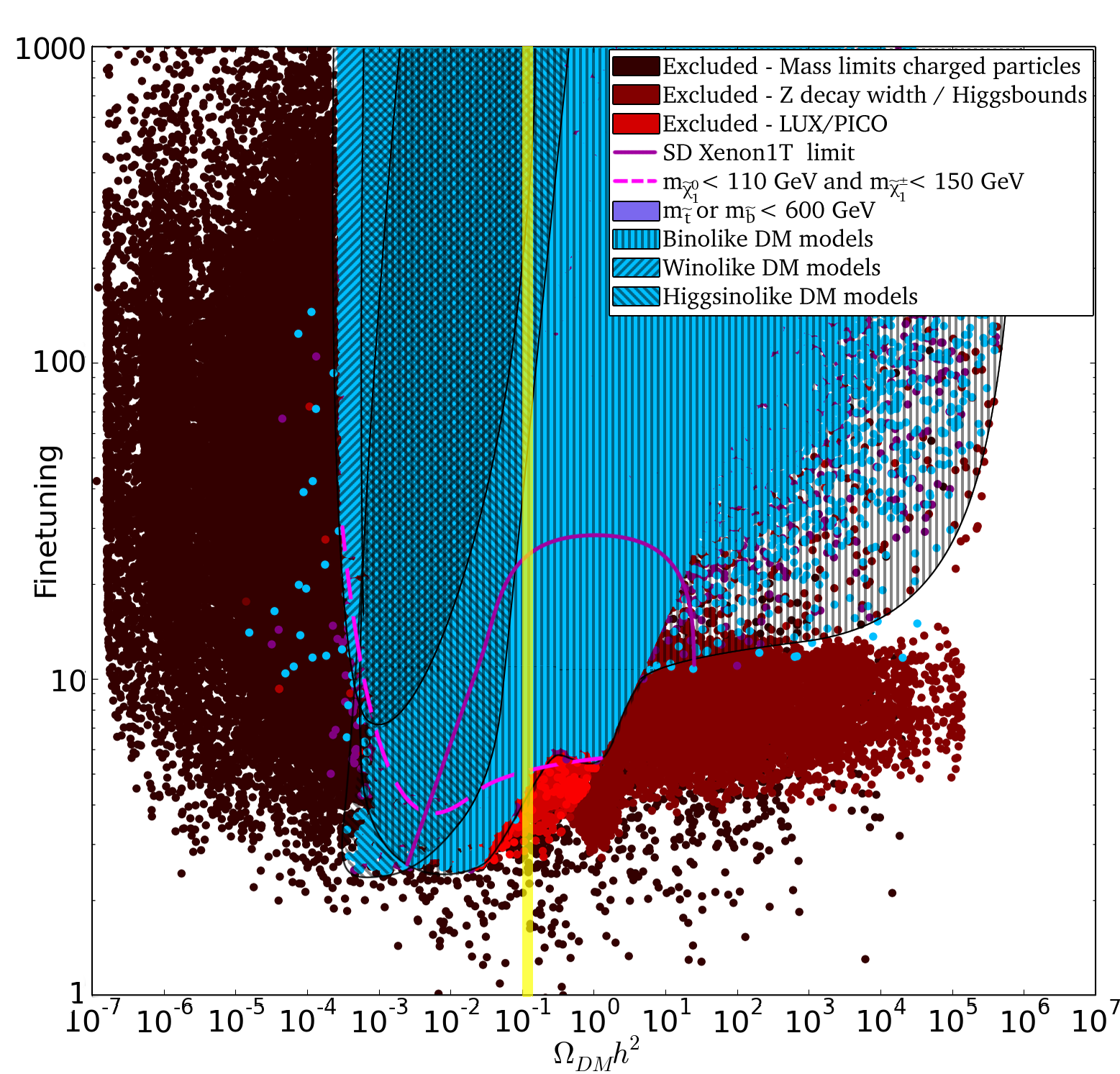}
	\caption{Fine-tuning as function of the DM relic density ($\Omega_{\rm DM} h^2$). Dark brown, maroon and red points indicate that the models are excluded due to mass limits on charged particles at LEP, bounds on the decay widths of the $Z$- and Higgs bosons, and LUX/PICO measurements on the WIMP-nucleus scattering cross section, respectively. The points indicated in purple are under pressure due to the LHC experiments that look for colored sparticles. In blue we show the allowed model points, with the corresponding DM composition indicated by the hatching. The lila solid curve indicates the predicted sensitivity from XENON1T \cite{Aprile:2015uzo} and the pink dashed curve indicates the predicted sensitivity from a proposed LHC search for bino-higgsino electroweakinos \cite{vanBeekveld:2016hbo}. The yellow band indicates $0.106 < \Omega_{\rm DM} h^2 < 0.130$. }
	\label{fig:FT}
\end{figure}

\noindent In figure \ref{fig:FT} the models with a FT less than 1000 are shown before and after applying limits. Before limits, the value for $\Omega_{\rm DM} h^2$ resulting from pMSSM models can range from $10^{-7}$ to $10^{6}$. Most models that have $\Omega_{\rm DM} h^2  < 10^{-4}$ are excluded due to the LEP limits on the chargino mass. These models have mostly light ($<100$ GeV) higgsino or wino DM particles, which are necessarily accompanied by a chargino of roughly the same mass. \\
For models that are not excluded we observe that a minimum FT value of 2.7 is obtained for $0.001 < \Omega_{\rm DM} h^2 < 0.05 $. Without using the minimalization algorithm, we would have obtained a minimal FT of approximately 10. For most of the low FT models, $C_{\mu}$ gives the dominant FT contribution. We have checked with Vevacious \cite{Camargo-Molina:2013qva,Lee2008,Wainwright:2011kj} that the points with lowest FT do not have a color/charge breaking minimum and have at least a metastable minimum that has a lifetime that exceeds that of our universe. Furthermore, none of the models are in tension with IceCUBE 2016 data \cite{Aartsen:2016exj, Scott:2012mq}. By evaluating the WIMP-nucleus scattering cross section, we predict that XENON1T \cite{Aprile:2015uzo} is sensitive to many of the low FT models. In case of non-discovery, this would increase the minimal FT of models that predict the right DM relic density to roughly 20, putting a natural version of the pMSSM in jeopardy.  \\
\begin{figure}[t]
	\centering
	\includegraphics[width=0.5\textwidth]{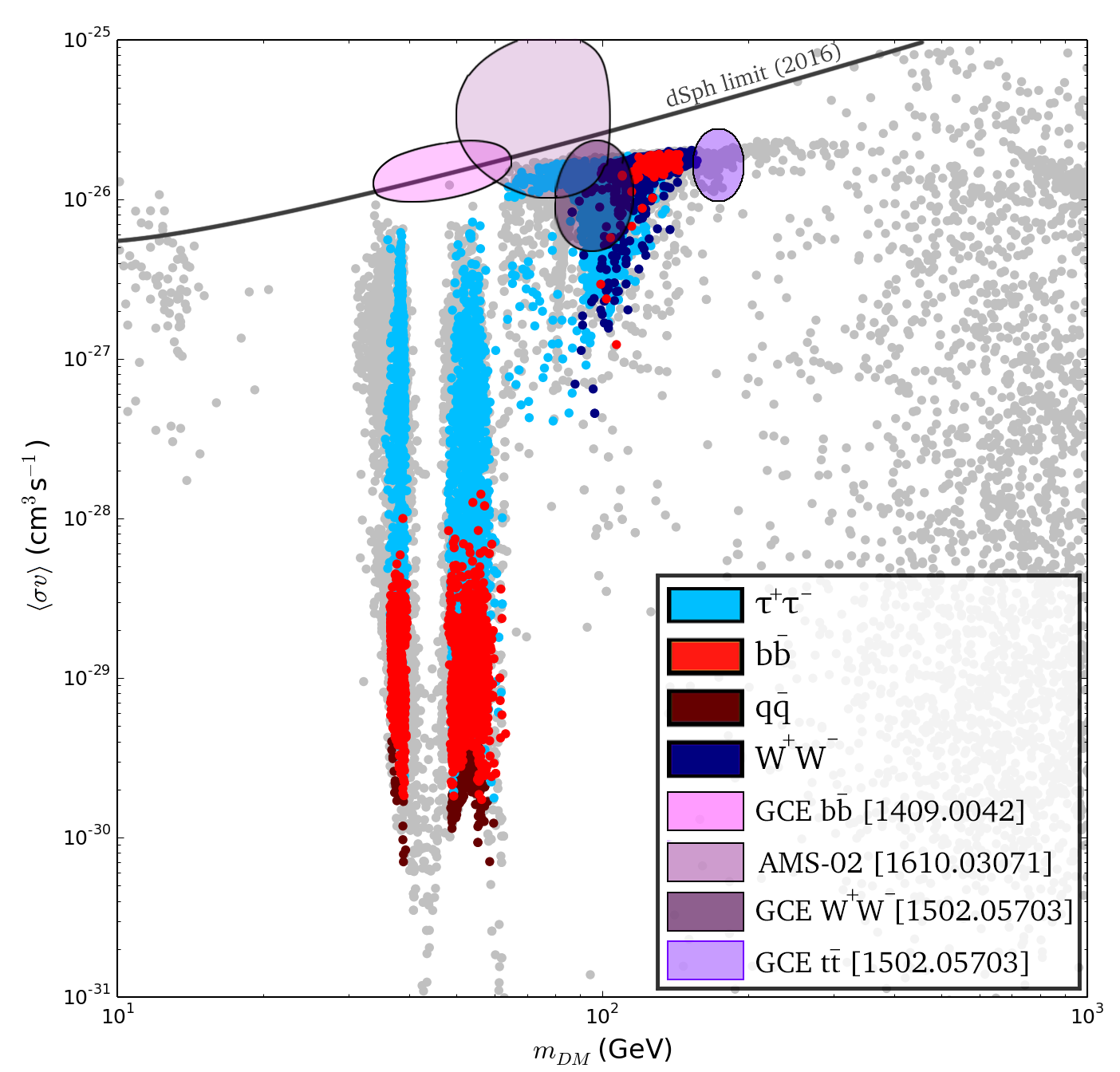}
	\caption{The present-day velocity-weighted dark matter annihilation cross section $\langle \sigma v \rangle$ (cm$^{3}$s$^{-1}$) as function of the dark matter mass $m_{\rm DM}$ (GeV) for models with FT $< 10$ (colored) and FT $ \geq 10$ (gray) and a relic density between $0.106 < \Omega_{\rm DM} h^2 < 0.130$. The color code indicates the dominant dark matter annihilation channel: dark blue for $W^+W^-$, light blue for $\tau^+ \tau^-$, red for $b\bar{b}$ and brown for $q\bar{q}$. Purple and pink shadings indicate the favored regions to explain the AMS-02 antiproton excess and the  Galactic Center photon excess. The dark gray line indicates the limit on the DM annihilation cross section derived from observations of dwarf galaxies assuming a 100$\%$ annihilation to $b\bar{b}$ \cite{Fermi-LAT:2016uux}.} 
	\label{fig:dom_chan}
\end{figure}

\begin{figure*}
	\centering 
 	\includegraphics[width=\textwidth]{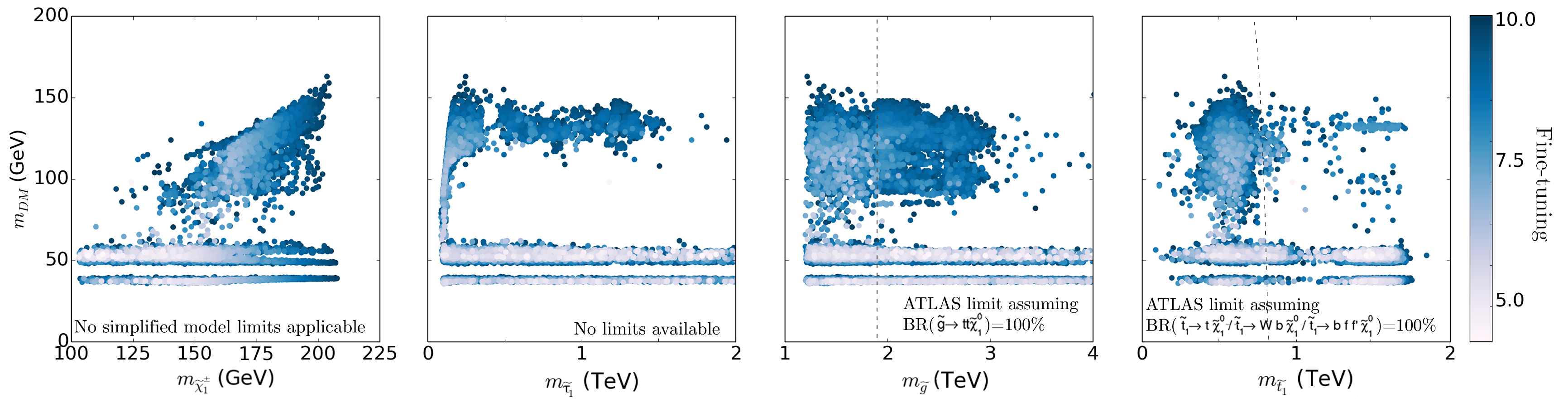}
	\caption{ Lightest chargino, stau, gluino and stop mass versus the DM mass for lowest-FT natural models satisfying all constraints (including the dark matter relic density). The FT is shown in color scale. The ATLAS 13 TeV search limits, produced using simplified SUSY models, are also shown for comparison. However, as explained in the text, these limits actually are not applicable to the majority of our models \cite{ATLAS:2017cxl, ATLAS-CONF-2017-021, ATLAS-CONF-2017-034, ATLAS-CONF-2017-037}.} 
    \label{fig:vars}
\end{figure*}

\begin{figure*}
	\centering
 	\includegraphics[width=\textwidth]{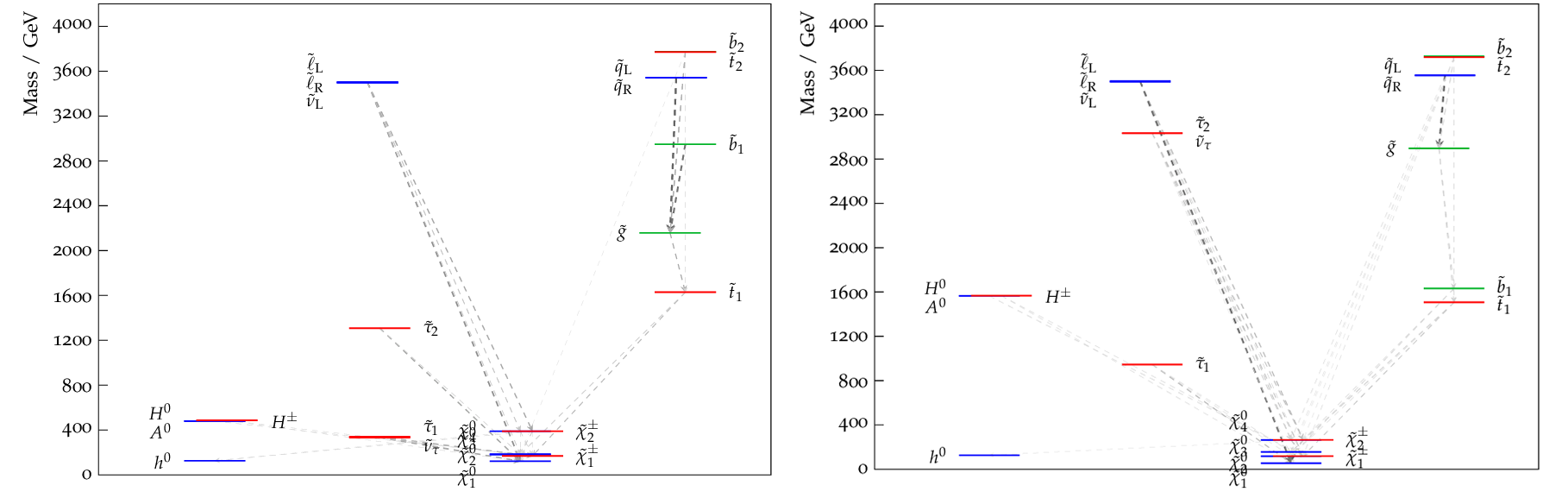}
	\caption{Characteristic mass spectrum for two of our low fine-tuning solutions with the correct relic density. All decays with a branching ratio larger than 10$\%$ are indicated by arrows. The decay arrows are plotted with a thickness and color related to the branching ratio (darker represents a higher branching ratio). The figure has been made using \textsc{PySLHA} \cite{Buckley:2013jua}.  }
    \label{fig:spectrum}
\end{figure*}

\begin{figure*}

	\centering
 	\includegraphics[width=\textwidth]{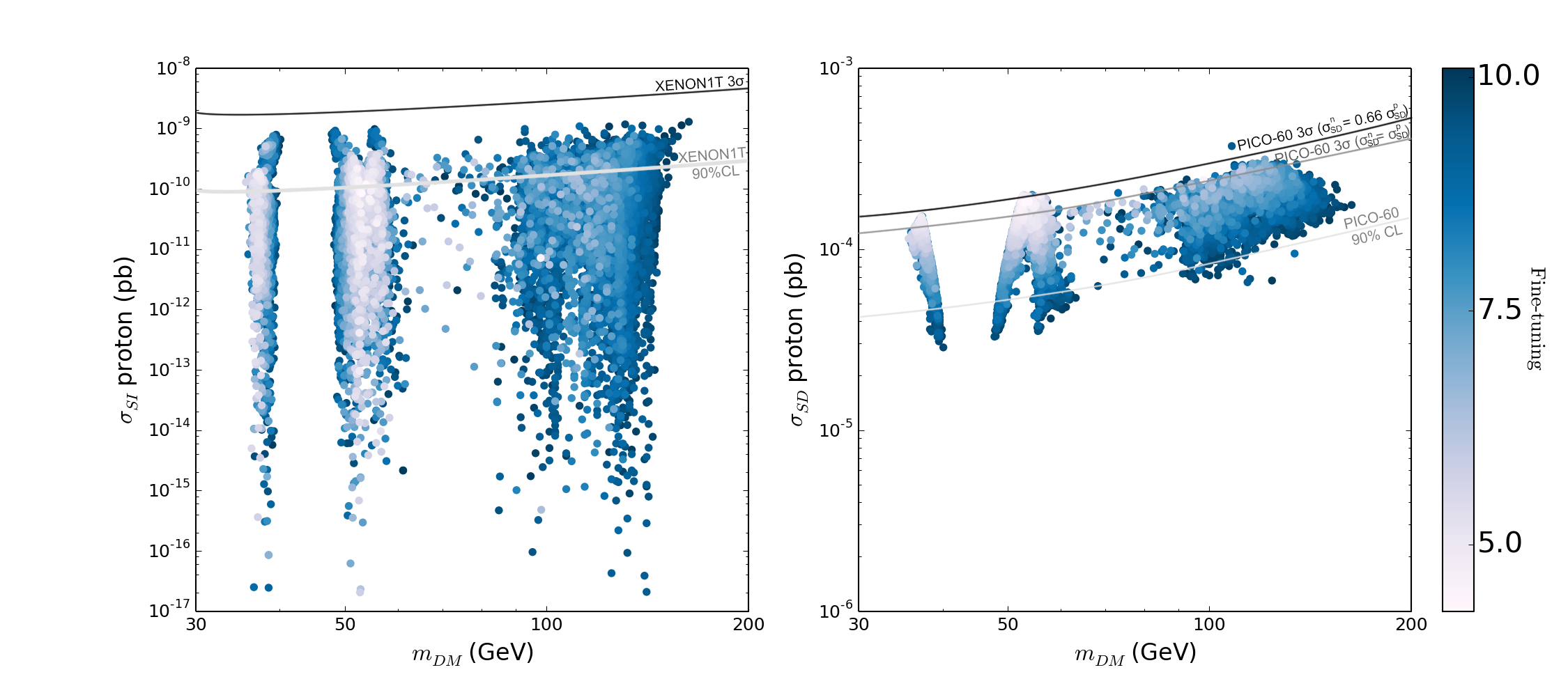}
      \caption{The spin-independent (left) and spin-dependent (right) WIMP-proton cross section versus the dark matter mass for the lowest-FT natural models satisfying all constraints  (including the dark matter relic density). The FT is shown in color scale. The $\sigma_{SI,p}$ XENON1T limit \cite{Aprile:2017iyp} and the $\sigma_{SD,p}$  PICO-60 limit \cite{Amole:2017dex} are also shown for comparison. Note that these experiments assume that the neutron-WIMP and proton-WIMP cross sections are equal to derive these limits. In our models, $\sigma_{SD, n}$ is always lower than $\sigma_{SD, p}$. To illustrate the effect of this, we show in the $\sigma_{SD,p}$ plot also the $3\sigma$ PICO limit for the assumption that $\sigma_{SD,n}/\sigma_{SD,p} = 2/3$. }
    \label{fig:sigmasisd}
\end{figure*}

\noindent The lowest FT models dominantly have a DM particle that is higgsino-like. It is well known that higgsino (and wino) DM with a mass around 100 GeV provides a too effective DM annihilation, resulting in $\Omega_{\rm DM} h^2$ well below the measured value. In addition, non-excluded models with a wino-like DM particle can only have a FT of larger than 6.5. If the lightest neutralino is a natural DM candidate, we also expect it to make up the entire DM relic density if the thermal freeze-out model is correct. We therefore demand $0.106<\Omega_{\rm DM} h^2<0.130$ and FT~$<10$, resulting solely in models with a bino($70-95\%$)-higgsino($30-5\%$) DM particle. A fraction of these points is excluded by LUX direct detection experiments, increasing the minimal value for FT to roughly 4.7. Figure \ref{fig:dom_chan} shows $\langle\sigma v\rangle$ versus the DM mass for these natural models, with the dominant DM annihilation channel indicated in color. We can distinguish three mass ranges for the DM particles: 35-40~GeV, 45-65~GeV and 80-155~GeV.\\ 

The first mass range (35-40~GeV) contains natural models with values for $\langle\sigma v\rangle$ that are orders of magnitude lower than the value $\langle \sigma v \rangle \simeq 3 \times 10^{-26}$ cm$^{3}$s$^{-1}$ that is typically predicted in simplified models for a thermal relic particle with a mass around 100~GeV. In the early universe the thermal 35-40~GeV DM particles annihilated via their higgsino component through an almost on-resonance s-channel exchange of a $Z$-boson, resulting in a lower DM relic density than is expected from $\langle \sigma v \rangle$ alone. The models with a DM particle in this mass range that have a light stau ($m_{\tilde{\tau}_1}~<200$~GeV) additionally annihilate through a t-channel stau exchange and therefore have significantly larger values for $\langle\sigma v \rangle$. These DM particles have a slightly lower higgsino component, causing the s-channel $Z$-boson annihilation in the early universe to be less efficient. Getting closer to $m_{\rm DM}=m_Z/2$ we find no solutions, as the annihilation is too efficient for low FT models, which all have a DM particle with a significant higgsino component. In the second mass range, we observe similar features in the vicinity of $m_{\rm DM} \simeq m_{h_0}/2$, only then caused by the s-channel exchange of a Higgs boson.
 \\
In the mass range of 80-155 GeV, three annihilation modes dominate the natural models: annihilation to $\tau^+\tau^-$ (via t-channel $\tilde{\tau}_1$ exchange), to $W^+ W^-$ (via t-channel $\tilde{\chi}^{\pm}_1$ exchange) and to $b\bar{b}$ (via t-channel $\tilde{b}_1$ exchange). Due to the mass of the DM particle that is necessarily higher, models where the DM particles annihilate to top pairs have slightly higher FT values of 13-19.  

\noindent None of these natural low FT points are in tension with limits obtained from dwarf galaxies \cite{Ackermann:2015zua, Ahnen:2016qkx, Fermi-LAT:2016uux}. Remarkably, some of the obtained models yield values for $\langle \sigma v \rangle$  that are in the range for explaining the Galatic Center (GC) photon excess \cite{Goodenough:2009gk, Caron:2015wda, TheFermi-LAT:2015kwa}, the excesses observed in dwarf galaxies \cite{Achterbeg:2015dca, Fermi-LAT:2016uux} and the AMS-02 antiproton excess \cite{Cuoco:2016eej}. In the case of our lowest FT natural models most likely only a fraction of the excesses seen in the GC would be due to DM annihilation. This motivates a further investigation of these excesses with a mixed DM and background explanation. \\
\noindent Most of these natural solutions are not in tension with recent LHC results, in spite of the presence of light sparticles (see figure \ref{fig:vars}). We find that, contrary to what is commonly assumed, we do not need a very low ($\leq 600 $~GeV) stop mass, sbottom mass or gluino mass to get low FT values, which is consistent with the findings in refs. \cite{Baer:2012cf, PhysRevD.87.115028dd, Casas:2014eca, Boehm:2013gst}. The stops start contributing substantially to the FT when the lightest stop is heavier than 2 TeV, while the ATLAS and CMS mass limits go up to only 850 GeV in the most optimistic scenario \cite{ATLAS:2016jaa,CMS:2016kcq}. This motivates LHC searches that look beyond the production of colored sparticles. To efficiently probe the natural low FT models, the LHC would need a dedicated low-mass stau search or a compressed chargino-neutralino search \cite{vanBeekveld:2016hbo}. These searches are complicated due to the low production cross section for staus and higgsino-like charginos/neutralinos and due to the presence of high background rates in case of the stau search.  \\ 
\noindent Two characteristic mass spectra for our lowest FT solutions are shown in figure \ref{fig:spectrum}. Decays for which the branching ratio is $>10\%$ are also shown. These figures illustrate why simplified model limits that ATLAS and CMS produce are not applicable to many of our models. Many sparticles in our models have a complicated decay chain, which would significantly reduce the amount of events in the signal region that the experiments specify, leading to a greatly reduced sensitivity. In the case of the electroweakinos, the composition of the particles also plays a role. The simplified-model limits given by the experiments are based on the assumption of a pure wino $\tilde{\chi}^{\pm}_1$ and $\tilde{\chi}^0_2$, while in our models we have a higgsino $\tilde{\chi}^{\pm}_1$ and a bino-higgsino $\tilde{\chi}^0_2$. The cross section for a higgsino chargino-neutralino pair is smaller compared to the cross section for a wino chargino-neutralino pair of the same mass. \\
\noindent The impact of direct detection experiments on the natural dark matter models can be fully attributed to the sensitivity on the spin-dependent cross section (see figure \ref{fig:sigmasisd}). The spin-independent WIMP-nucleon cross section for these natural dark matter models spans a large range of values ($10^{-9}-10^{-17}$~pb). The spin-dependent WIMP-nucleon cross section is much more constrained. This is directly related to the higgsino component in the dark matter particle: a higher higgsino component increases the $Z \tilde{\chi}^0_1\tilde{\chi}^0_1$ coupling, thereby increasing the spin-dependent WIMP-nucleon cross section. The value for $\mu$ has to increase in order to reduce the spin-dependent WIMP nucleon cross section, which causes the FT to increase as well. 
\section{Conclusions}
\noindent In this paper we minimized the fine-tuning of the pMSSM, taking into account all experimental constraints. Based on naturalness arguments (i.e. demanding FT $<10$) on the $Z$-boson mass combined with demanding the observed DM relic density, we predict a DM particle that is bino-higgsino-like with a mass of 35-155~GeV as most natural SUSY DM candidate. The LUX experiment has already been able to cut into the space of low FT models, increasing the minimal FT from 2.7 to 4.7. Remarkably, the natural low FT models are not under pressure by LHC searches for stops, as stops start contributing substantially to the FT when $m_{\tilde{t}_1}>2$~TeV, while the stop searches place limits of $m_{\tilde{t}_1}>850$~GeV in the most optimistic scenarios \cite{ATLAS:2016jaa,CMS:2016kcq}. Interestingly, some of the lowest-FT natural solutions are consistent with the SUSY dark matter explanations for various anomalies observed in astrophysical experiments  \cite{FermiLAT:2012aa,Caron:2015wda, Fermi-LAT:2016uux, Achterbeg:2015dca, Cuoco:2016eej, Cui:2016ppb}. Direct detection experiments and the dedicated LHC searches will be able to test this region of natural models within the next five years.\\

\paragraph{Acknowledgments.}
M. van Beekveld, W. Beenakker and R. Peeters acknowledge support by the Foundation for Fundamental Research of Matter (FOM), programme 156, "Higgs as Probe and Portal". The work of R. Ruiz de Austri was supported by the Ramon y Cajal program of the Spanish MINECO. He also acknowledges the support of the grants FPA2014-57816-P and FPA2013-44773, the Severo Ochoa MINECO project SEV-2014-0398 and especially the support of the Spanish MINECO's Consolider-Ingenio 2010 Programme under grant MultiDark CSD2009-00064. Furthermore the authors would like to thank H. Baer for his helpful suggestions and comments during the writing of our in-house FT code.

%
\bibliography{main}
\end{document}